\documentclass[a4paper,fleqn,usenatbib]{mnras}

\usepackage{newtxtext,newtxmath}

\usepackage[T1]{fontenc}
\usepackage{ae,aecompl}

\usepackage{graphicx}	
\usepackage{amsmath}	
\usepackage{amssymb}	

\title[A photometric parallax distance to V2051\,Oph]
      {Infrared photometry of the dwarf nova V2051 Ophiuchi: \\
       I - The mass donor star and the distance \thanks{Based on observations
       obtained at the Southern Astrophysical Research (SOAR) telescope, which
       is a joint project of the Minist\'{e}rio da Ci\^{e}ncia, Tecnologia,
       Inova\c{c}\~{a}os e Comunica\c{c}\~{a}oes (MCTIC) do Brasil, the U.S.
       National Optical Astronomy Observatory (NOAO), the University of North
       Carolina at Chapel Hill (UNC), and Michigan State University (MSU).} }

\author[Wojcikiewicz, Baptista \& Ribeiro]{
Eduardo Wojcikiewicz,$^{1}$
Raymundo Baptista,$^{1}$
Tiago Ribeiro $^{2}$
\\
$^{1}$Departamento de F\'{i}sica, Universidade Federal de Santa Catarina,
 Campus Trindade, 88040-900 Florian\'{o}polis, SC, Brazil\\
 $^{2}$Departmento de F\'{i}sica, Universidade Federal de Sergipe, Jardim Rosa
 Elze, 49100-000 S\~{a}o Crist\'{o}v\~{a}o, SE, Brazil
}

\date{Accepted XXX. Received YYY; in original form ZZZ}

\pubyear{2018}

\begin{document}
\label{firstpage}
\pagerange{\pageref{firstpage}--\pageref{lastpage}}
\maketitle

\begin{abstract}
We report the analysis of time-series of infrared $JHK_s$ photometry of the
dwarf nova V2051\,Oph in quiescence. We modelled the ellipsoidal variations
caused by the distorted mass-donor star to infer its $JHK_s$ fluxes. From
its infrared colors we estimate a spectral type of $M(8.0\pm 1.5)$ and an
equivalent blackbody temperature of $T_\mathrm{BB}=(2700\pm270)\,K$. We used
the Barnes \& Evans relation to infer a photometric parallax distance of
$d_\mathrm{BE}=(102\pm16)$ pc to the binary. At this short distance, the
corresponding accretion disc temperatures in outburst are too low to be
explained by the disc-instability model for dwarf nova outbursts,
underscoring a previous suggestion that the outbursts of this binary are
powered by mass-transfer bursts.
\end{abstract}




\section{Introduction}

Dwarf novae are short period interacting binaries in which an evolved, late
type star overfills its Roche lobe and transfers matter to a companion white
dwarf via an accretion disc. These systems show recurrent outbursts in which
the disc brightens for $2-5$ mag on timescales of days to weeks. Outbursts
are thought to be driven either by a sudden increase in mass transfer rate
\citep[Mass Transfer Instability Model, MTIM, e.g.,][]{Bathpringle1981} or by
a thermal-viscous disc instability \citep[Disc Instability Model, DIM, e.g.][]
{Lasota2001}. MTIM interprets the outburst as the response of a high viscosity
disc ($\alpha \sim 0.1-1.0$) to a burst of enhanced mass transfer from the
mass-donor star.
DIM assumes matter is fed at a roughly constant rate to a low viscosity disc,
progressively raising its surface density and gas temperature until it
surpasses a critical limit, $T_\mathrm{crit1}\simeq6000\,K$, where a
thermal-viscous instability sets in, greatly enhancing the local viscosity
parameter ($\alpha_\mathrm{hot}\simeq10\,\alpha_\mathrm{cool}\sim0.1$). The
instability propagates as a heating wave, bringing the whole disc into a
high-viscosity regime which enables the fast accretion of the accumulated gas
\citep[e.g.,][]{Lasota2001}. A disc annulus switches back to quiescence (and
to the low viscosity regime) when its temperature falls below $T_\mathrm{crit2}
\sim(10000-7000)\,K$. This limit-cycle scheme implies that the disc
temperatures must be $T<T_\mathrm{crit1}$ in quiescence and $T>T_\mathrm{crit2}$
during outbursts. Therefore, the comparison of observed outburst disc
temperatures with $T_\mathrm{crit1},T_\mathrm{crit2}$ provides a key test for DIM.

V2051 Oph is an ultra short-period eclipsing dwarf nova (orbital period
$P_\mathrm{orb}=90$ min) discovered by \citet{Sanduleak72}. \citet{Warner83} and
\citet{Cook83} reported optical light curves showing large-amplitude
($\gtrsim30\%$) flickering (random brightness variations of $0.1-1$ mag), deep
eclipses ($\Delta M_\mathrm{B}\simeq2.5\;\text{mag}$) and a plethora of different
eclipse profiles.
Superoutbursts were observed and
superhumps were detected by \citet{Kiyota98} and \citet{Vrielmann03}, implying
that V2051 Oph is an SU UMa type dwarf nova. \citet{Baptistaetal98} used HST
and ground-based observations to constrain the binary parameters finding a
mass ratio of $q = 0.19\pm 0.03$, an inclination of $i=83.3^\circ\pm1.4^\circ$,
star masses and radii of $M_1=(0.78\pm0.06)\,M_\odot$, $M_2=(0.15\pm0.03)\,M_\odot$,
$R_1=(0.0103\pm0.0007)\,R_\odot$ and $R_2=(0.16\pm0.01)\,R_\odot$. \citet{Saito06}
modelled the extracted UV-optical white dwarf spectrum to find a distance
estimate of $92^{30}_{-35}\,pc$. \citet{Baptistaetal07} found that if V2051\,Oph
is closer than $120\,pc$, its outbursting disc is cooler than $T_\mathrm{crit2}$
everywhere, therefore excluding DIM as a viable explanation for its outbursts.

Here we model the ellipsoidal modulation seen in infrared light curves of
V2051\,Oph to extract the fluxes of its mass-donor star and to infer a
photometric parallax distance to the binary. This paper is organized as follows.
Sect.\,2 describes the data reduction procedures, while Sect.\,3 presents the
data analysis and the results. The results are discussed in Sect. 4 and
summarized in Sect. 5.

\section{Observations and data reduction}

We used the OSIRIS Infrared Imager and Spectrograph attached to the 4.1\,m
SOAR Telescope at Cerro Pach\'{o}n, Chile, to collect time-series of $JHK_s$
photometry of V2051\,Oph in 2013 June 20th, while the star was in quiescence.
The observations are summarized in Table ~\ref{tab:observations}. The fourth
column gives the number of exposures in each passband, while the fifth column
lists the binary cycles (E) covered by each run. We adopted a single exposure
time of $10\,s$ for all runs and employed a 2x3 dithering pattern with an offset
of $10''$ between positions. We observed two full orbital cycles in the $K_S$
band and about 1.5 orbital cycles in the $H$ and $J$ bands. The images were
obtained with clear, photometric skies and the full moon near the field, with
seeing ranging $1''-1.8''$.

\begin{table}
 \centering
  \caption{Journal of observations of V2051 Oph. The binary cycle $E$ is calculated with respect to the ephemeris of Eq.~(\ref{eq:linear-ephem}).}
  \label{tab:observations}
  \begin{tabular}{ccccc} 
    \hline
    Band & $UT_\mathrm{start}$ & $UT_\mathrm{end}$ & $N_\mathrm{exp}$ &  E  \\
    \hline
    $K_S$ & 23:43 & 2:47 & 214 & 211741-211742\\
    $H$ & 2:58 & 5:46  & 181 & 211743-211744\\
    $J$ & 6:10 & 8:46  & 180 & 211745-211746\\
    \hline
  \end{tabular}
\end{table}

The data were reduced with the IRAF package \footnote{IRAF is distributed by
  the National Optical Astronomy Observatories, which are operated by the
  Association of Universities for Research in Astronomy, Inc., under
  cooperative agreement with the National Science Foundation.}.
Exposures were corrected from non-linearity effects of the instrument using
the third-order polynomial of \citet{Pogge1999}. Differences in pixel-to-pixel
sensitivity were compensated for dividing each corrected exposure by a
normalized flat-field frame. Since the sky provides a substantial contribution
to the IR light, we followed the standard procedure of dithering the telescope
into $n_\mathrm{dith}$ positions between exposures and median-combining the
resulting set of images to obtain an average sky frame. This median sky frame
was then subtracted from the corresponding images. Fluxes for a reference star,
the variable star, and a set of nearby comparison stars were extracted with
scripts using apphot/IRAF tasks in order to obtain light curves of differential
photometry. These light curves were flux calibrated using the 2MASS $JHK_s$
magnitudes of the reference star and the zero point constants of
\citet{Skrutskie2006}.

Exposure times were transformed from Universal Time (UT) to Terrestrial Time
(TT) with the correction $\Delta t_\mathrm{UT\mapsto TT}=-0.00078\,d$ and from
Heliocentric Julian Date (HJD) to Baricentric Julian Date (BJDD) with the
additional correction $\Delta t_\mathrm{HJD\mapsto BJDD}=-0.00079\,d$, adequate
for the epoch of the observations and the target coordinates, according to the
code by \citet{Stumpff1980}. The resulting light curves were phase-folded
according to the linear ephemeris of \citet{Baptistaetal2003},

\begin{equation}
  T_{mid}(BJDD) = 2443245.97752(3)+0.0624278634(3)\times E
  \label{eq:linear-ephem}
\end{equation}

\noindent where $T_\mathrm{mid}$ is the white dwarf mid-time eclipse and $E$ is
the binary cycle. V2051\,Oph shows cyclical period modulations
\citep{Baptistaetal2003}. The difference between the predicted eclipse timings
for the linear and sinusoidal ephemeris amounts to
$\Delta t_\mathrm{EFEM}=-0.00210\;d$ for the epoch of our observations. Combining
all these time corrections leads to a net phase shift of $\Delta\phi=0.0217$,
which centers the eclipse at phase zero.

During the observing night, a small $\sim0.1$ mag brightening was observed
along the $H$ band run followed by a $\sim 0.1$ mag decline during the $J$
band run. A cubic spline was fitted to the uneclipsed parts of these light
curves to level the out-of-eclipse average flux.

\section{Data analysis and results}

\begin{figure}
\centering
  \includegraphics[width=0.85\textwidth, angle = -90]{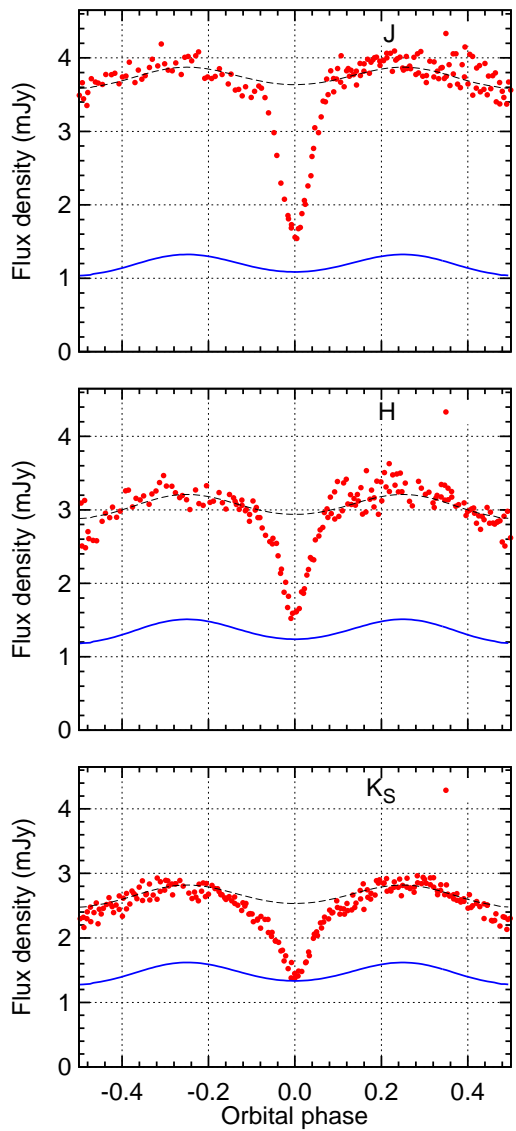}
   \caption{Phase-folded orbital $JHK_S$ light curves of V2051 Oph (dots),
     the modelled mass-donor star contribution (solid curve) and the ellipsoidal
     curve plus the constant disc flux $f_1$ (dashed curve).}
   \label{fig:curves}
\end{figure}

We modelled the ellipsoidal modulation caused by the changes in phase of the
projected area of the distorted mass-donor star with a computer code,
described in \citet{Ribeiroetal2007}. We assumed a mass ratio of
$q=(0.19\pm0.03)$, an inclination $i=83.3^o\pm1.4^o$ \citep{Baptistaetal98},
and that the mass-donor star has a uniform surface brightness. We adopted a
gravity darkening coefficient of $\beta=0.05$ \citep{Sarna89}, and the
non-linear square-root limb-darkening law of \citet{Diaz92},

\begin{equation}
  I_\nu = I_{0,\nu} \left[1-a_\nu\left(1-\cos\gamma\right)-b_\nu\left(1-\sqrt{\cos\gamma}\right)\right]
  \label{eq:limbdarkening}
\end{equation}

\noindent where the $a_\nu$ and $b_\nu$ coefficients are $a_J = -0.465$,
$b_J=1.199$, $a_H = -0.454$, $b_H=1.173$, $a_{K_S}=-0.448$ and $b_{K_S} =1.066$
\citep{Claret98}, $I_0$ is the undarkened intensity and $\gamma$ is the angle
between the normal to the surface element and the line of sight. The primary
($-0.1$ to $+0.1$) and secondary ($+0.4$ to $-0.4$) eclipse phases are removed
from the light curve and the remaining curve is input to the fitting code,
which returns the best-fit orbital contribution of the mass-donor star
(represented by its flux at phase zero, $f_2$) plus a constant flux level
(attributed to the accretion disc, $f_1$). The results of this fit are shown
in Fig.~\ref{fig:curves} and listed in Table~\ref{tab:fluxes}.

\begin{table}
  \centering
  \caption{Modelled fluxes for the mass-donor star ($f_2$) and the accretion disc ($f_1$)}
  \label{tab:fluxes}
  \begin{tabular}{ccc} 
   \hline
   Band  & $f_2$(mJy) & $f_1$(mJy) \\
   \hline
   $J$ & $1.09 \pm 0.17$   & $2.21\pm0.26$\\
   $H$ & $1.24 \pm 0.13$   & $1.74\pm0.13$\\
   $K_S$ & $1.33 \pm 0.14$ & $1.18\pm0.15$\\
   \hline
  \end{tabular}
\end{table}

\begin{figure}
  \centering
  \includegraphics[width=0.36\textwidth, angle = -90]{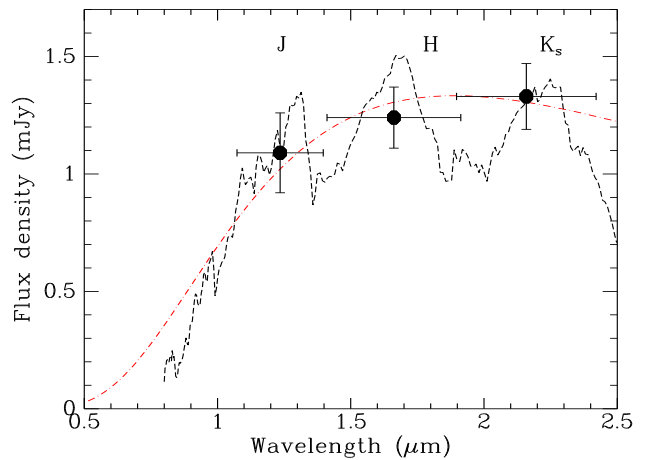}
    \caption{Mass-donor star $JHK_s$ fluxes (dots with error bars), best-fit
      $T_\mathrm{BB}=2700\;K$ blackbody model (dot-dashed line), and scaled
      spectrum of an M8 star (dashed line) from \citet{Testi09}.}
    \label{fig:bb}
\end{figure}

The amplitude of the modulation, the depth of the secondary eclipse (at phase
$\pm0.5$) and the contribution of the mass-donor star increase with wavelength,
indicating that it has a very late-type spectrum. On the other hand, the
primary eclipse depth and the disc flux $f_1$ decrease with increasing
wavelength, telling us that the accretion disc is bluer than the mass-donor star.
A blackbody fit to the mass-donor star fluxes yields a temperature of
$T_\mathrm{BB}=\left(2700\pm270\right)\;\text{K}$ and a preliminary distance
estimate of $d_\mathrm{BB}=\left(107\pm17\right)\;\text{pc}$. The inferred
$JHK_s$ fluxes of the mass-donor star and the best-fit blackbody to these
fluxes are shown in Fig.~\ref{fig:bb}. The predicted optical flux of the
mass-donor star ($\leq 0.1$ mJy) is consistent with the remaining flux at
mid-eclipse observed by \citet{Baptistaetal98}.

The extracted $JHK_s$ fluxes were converted to magnitudes in order to infer
the infrared colours of the mass-donor star (Table~\ref{tab:magnitudes}).
The inferred colours allow us to estimate both the spectral type and the
distance to the system by the method of \citet{Bailey81} using the
empirical Barnes-Evans relations for the infrared provided by
\citet{Beuermann06}. Since the measured colours are consistent with those of
late $M$ type stars, we adopted the lower main sequence ($M6.5 \leqslant
SpT \leqslant L8$) empirical relation of \citet{Beuermann06} to obtain the
$K$ band surface brightness of a star as a function of its spectral type,

\begin{equation}
  S_K = 9.651 - 0.88541X + 0.068535X^2 - 0.00211177X^3
  \label{eq:surfaceflux}
\end{equation}

\noindent where the parameter X for $M$ stars is given by $SpT =
M\left(20-X\right)$, $11 \leqslant X \leqslant 20$. Once $S_K$ is known, the
distance can be estimated from the Barnes-Evans relation \citep{BarnesEvans1978},

\begin{equation} 
  S_K = m_K + 5\log\left(R/R_\odot\right)-5\log\left(d/10\;\text{pc}\right) \, .
  \label{eq:barnesevans}
\end{equation}

In order to estimate the spectral type of the mass-donor star, we applied
the equations of \citet{Eliasetal83} to transform the measured magnitude and
colours from the 2MASS to the CIT photometric system and matched the
resulting magnitude and colours to the sample of cool late type dwarfs of
\citet{Cruz2003}, taking into account the larger size of the mass-donor star
with respect to isolated stars of same mass, to find $SpT = M(8.0\pm1.5)$.
We then performed Monte Carlo simulations, applying random gaussian noise to
the spectral type, the $K_s$ band magnitude, and the radius of the mass-donor
star \citep[$R_2=0.16\pm0.01\,R_\odot$,][]{Baptistaetal98} to
Eqs.~(\ref{eq:surfaceflux}) and (\ref{eq:barnesevans}) to obtain
$S_K=(5.24\pm0.21)$ and a photometric parallax distance estimate of
$d_\mathrm{BE}=(102\pm16)\,pc$.

\begin{table}
	\centering
	\caption{Magnitudes and colours of the mass-donor star}
	\label{tab:magnitudes}
	\begin{tabular}{cc|cc} 
		\hline
		Band  & (mag) & Colour & (mag)  \\
		\hline
		$J$ & $15.42 \pm 0.10$ & $J-H$ & $0.62 \pm 0.11$  \\
		$H$ & $14.79 \pm 0.05$ & $H-K_S$ & $0.55 \pm 0.08$  \\
		$K_S$ & $14.25 \pm 0.05$ & $J-K_S$ & $1.17 \pm 0.11$  \\
		\hline
	\end{tabular}
\end{table}

According to \citet{Luhman03}, a typical $M8$ star has $T_{eff}=2710K$, in
good agreement with our estimate. Our result is also consistent with the
spectral type $SpT = M(7\pm1)$ obtained by \citet{Hamiltonetal11}, and the
$T_\mathrm{eff}=2600\,K$ and $SpT=M7$ expected for a dwarf nova with
$P_\mathrm{orb}=90$\,min \citep{Knigge06}. The scaled spectrum of the M8 star
LP\,412-31 \citep{Testi09} is shown in Fig.\,\ref{fig:bb} for illustration
purposes.

The distance estimates to V2051\,Oph obtained independently from the fit to
the white dwarf UV-optical spectrum \citep[$d=92_{-35}^{+30}\,pc$,][]{Saito06}
and from the ellipsoidal modulation of the mass-donor star in the infrared
($d_{BE}=102\pm16\,pc$) are consistent with each other at the 1-$\sigma$
confidence level.

\section{Discussion}

We investigated whether our results could be affected by interstellar
extinction or by irradiation effects on the mass-donor star from a hot
white dwarf or accretion disc.

The signature of irradiation is the presence of an orbital
hump centred at phase $\pm 0.5$, when the irradiated face of the mass-donor
star is seen face-on. \citet{Saito06} measured a white dwarf temperature of
$T_\mathrm{WD}= 9500^{+2900}_{-1900}\,K$. At such low white dwarf temperature,
there are not enough UV photons to produce detectable irradiation effects.
Furthermore, at a distance of about $100\,pc$, even in outburst the accretion
disc of V2051\,Oph is everywhere cooler than $\sim10000\,K$ and cannot lead to
significant irradiation effects. We confirmed that expectation by modifying the
ellipsoidal variation code to allow modelling of an additional bright spot on
the inner face of the mass-donor star caused by possible irradiation effects.
The resulting intensity of this additional spot is negligible in all bands,
in agreement with the lack of evidence for any orbital hump at phase $\pm 0.5$.
 
Spectroscopy of V2051\,Oph shows no evidence of the 2200\,\AA\ absorption
feature \citep{Wattsetal86, Baptistaetal98} or the Na\,I $\lambda 5890,5896$
absorption doublet \citep{Steeghsetal01}, suggesting that interstellar
extinction effects are also negligible. Indeed, from \citet{Schlafly11} we
estimate $E(B-V)\leq 0.42$\,mag kpc$^{-1}$ towards the direction of V2051\,Oph,
which leads to reddening corrections of $\Delta K_s\leq 0.013$\,mag and
$\Delta (J-K_s) \leq 0.025$\,mag for a distance of 102\,pc, small in
comparison to the uncertainties in the derived magnitude and colours.

The error in the inferred photometric parallax distance estimate is largely
dominated by the uncertainties in the measured IR magnitude and colours and
by the uncertainties in the spectral classification of very late-type stars
\citep[e.g.,][]{Testi09}, with non significant influence from irradiation or
interstellar extinction effects.

The distance to the binary is a key factor to distinguish which of the two
available models is responsible for the outbursts of V2051 Oph. Because the
DIM thermal limit-cycle relies on the partial ionization of hydrogen, disc
temperatures are strongly constrained in the DIM framework, with quiescent
discs bound to $T<T_\mathrm{crit1}\simeq6000\,K$ and hotter outbursting discs
with $T>T_\mathrm{crit2}=(10000-7000)\,K$ for $R\simeq(0.02-0.5)\,R_\odot$. There
are no temperature restrictions for discs in the MTIM. Disc surface brightness
derived via eclipse mapping techniques can be transformed to brightness
temperatures (under the assumption of optically thick disc emission) if the
distance to the binary is known \citep{Baptista2016}. For a given surface
brightness map, shorter distances imply fainter and cooler discs.
\citet{Baptistaetal07} found that, during outbursts, the V2051 Oph disc
brightness temperatures remain everywhere below $T_\mathrm{crit2}$ if the
distance is lower than $120\,pc$. Our photometric parallax distance estimate
indicates it is not possible to explain the outbursts of this dwarf nova
within the DIM framework, which leaves MTIM as the only current alternative
explanation for the outbursts of V2051 Oph.

\section{summary}

We modelled the ellipsoidal modulation caused by the distorted mass-donor
star to infer its fluxes in the $JHK_S$ bands and used those fluxes to obtain
its magnitudes and colours. The derived colours matches those of a typical
lower main-sequence star of spectral type $M(8.0\pm1.5)$, with temperatures
in the $T_\mathrm{eff}=2700\,K$ range, in good agreement with the spectral
type inferred by \citet{Hamiltonetal11} and that expected for a dwarf nova
with similar orbital period \citep{Knigge06}.

The infrared surface brightness and colours imply a photometric parallax
distance estimate of $d_\mathrm{BE}=(102\pm 16)\,pc$ for V2051 Oph, consistent
at the 1-$\sigma$ confidence level with the independent estimate by
\citet{Saito06}. At this short distance, the outbursts of V2051 Oph occur at
disc temperatures everywhere below the minimum outburst temperature required
by the thermal-viscous disc instability model and is, therefore, incompatible
with this dwarf nova outburst model. This underscores the previous suggestion
that the outbursts of this dwarf nova are powered by bursts of mass transfer
from its donor star.

\section*{Acknowledgements}

E.\ W.\ acknowledges financial support from CAPES (Brazil) and CNPq (Brazil).
R.\ B.\ acknowledges CNPq grant 308.946/2011-1.



\bibliographystyle{mnras}
\bibliography{bibliography} 

\bsp	
\label{lastpage}
\end{document}